\documentclass[%
 reprint,
superscriptaddress,
 amsmath,amssymb,
 aps,
pra
]{revtex4-2}

\usepackage{xcolor}
\usepackage{graphicx}
\usepackage{dcolumn}
\usepackage{float}
\usepackage{bm}

\usepackage{svg}
\usepackage{subfiles} 
\usepackage[colorlinks= true]{hyperref}

\begin{document}

\title[Exceptional point sensing via energy loss profile in a non-Hermitian system]{Exceptional point sensing via energy loss profile in a non-Hermitian system}

\author{Parul Sharma}
\author{Brijesh Kumar}
\author{Nihar Ranjan Sahoo}\affiliation{Laboratory of Optics of Quantum Materials, Department of Physics, IIT Bombay, Mumbai - 400076, India}
\author{Anshuman Kumar}
\email{anshuman.kumar@iitb.ac.in}
\affiliation{Laboratory of Optics of Quantum Materials, Department of Physics, IIT Bombay, Mumbai - 400076, India}

\begin{abstract}
The heightened sensitivity observed in non-Hermitian systems at exceptional points (EPs) has garnered significant attention. Typical EP sensor implementations rely on precise measurements of spectra and importantly, for real time sensing measurements, the EP condition ceases to hold as the perturbation increases over time, thereby preventing the use of high sensitivity at the EP point. In this work, we present an new approach to EP sensing which goes beyond these two traditional constraints. Firstly, instead of measuring the spectra, our scheme of EP based sensing is based on the observation of decay length of the optical mode in finite size gratings, which is validated via coupled mode theory as well as full wave electrodynamic simulations. Secondly, for larger perturbation strengths, the EP is spectrally shifted instead of being destroyed -- this spectral shift of the EP is calibrated and using this look-up table, we propose continuous real time detection by varying the excitation laser wavelength. As a proof of principle of our technique, we present an application to the sensing of coronavirus particles, which shows unprecedented limit of detection. These findings will contribute to the expanding field of exceptional point based sensing technologies for real time applications beyond spectral measurements.

\end{abstract}

\maketitle

\emph{Introduction--}
Exceptional points (EPs) have gained prominence as an effective tool for manipulating the behaviour of parity time symmetric non Hermitian systems\cite{Li2023,zdemir2019, Miri2019}. In contrast to conservative systems that obey conservation laws, such open systems interact and exchange energy with their surrounding environment\cite{Bender1999, Bender2007, ElGanainy2018, Feng2017}. EPs are unique points in the parameter space of such where multiple eigenfrequencies and eigenvectors converge simultaneously\cite{Zhang2018,Li2021,Song2021,PhysRevLett.123.193901}. In the vicinity of EP, the eigenfrequency spectrum of these systems undergoes a transition from real to complex-valued, giving rise to several exotic effects related to unidirectional invisibility\cite{Regensburger2012,Feng2013, PhysRevLett.106.213901}, lasing\cite{Hodaei2014,Lafalce2019,Brandstetter2014} and enhanced sensitivity\cite{Wiersig2020,Chen2017,Hodaei2017,PhysRevLett.130.227201}.

There are two characteristic features of typical implementations of EP sensors reported so far. Firstly, EP sensors are usually based on precise measurements of spectra, where the frequency splitting as a function of a perturbation is measured. Secondly, for real time sensing measurements, the EP condition ceases to hold as the strength of the perturbation increases over time, thereby preventing the use of high sensitivity at the EP point. In this work, we present an new approach to EP sensing which goes beyond these two traditional constraints. Firstly, instead of a spectral measurement, our unique scheme of EP based sensing is based on the observation of decay length of the optical mode, which is validated via coupled mode theory as well as full wave electrodynamic simulations. To the authors' knowledge there is no report on such a sensing approach in literature so far. Secondly, our sensor architecture maintains the EP condition for a wide range of perturbation strengths. For even larger deviations, the EP is spectrally shifted instead of being destroyed -- this spectral shift of the EP is calibrated and using a ``look-up table" approach, we propose continuous real time detection by varying the excitation laser wavelength. As a proof of principle of our technique, we present an application to the sensing of coronavirus particles, which shows unprecedented limit of detection up to a single virus particle.

\begin{figure}[ht]
  \centering
  \includegraphics[width =\columnwidth]{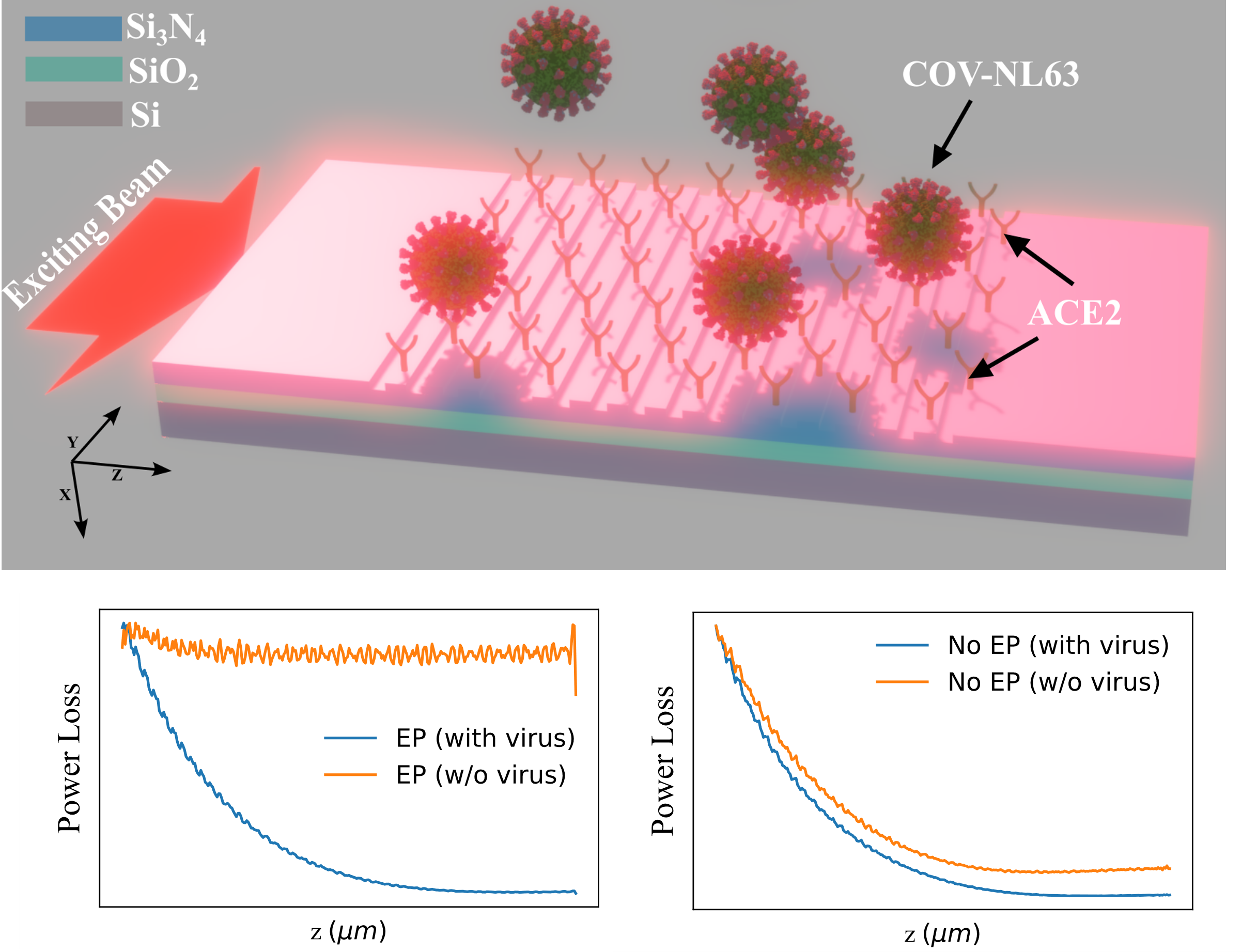}
   \caption{Schematic diagram for detecting a single coronavirus particle (CoV-NL63) and RBD spike protein using silicon nitride subwavelength gratings functionalized with human angiotensin-converting enzyme 2 protein (ACE2). The laser is coupled in-plane perpendicular to the grating and the far field spatial profile of the radiated intensity is measured at EP and non-EP duty cycles with virus and without virus respectively.}
  \label{fig1}
\end{figure}

\emph{Sensor architecture and analysis--}
In our study, we investigated exceptional points (EPs) in a geometry consisting of a periodic array of rib waveguides forming a grating. The architecture including the materials is shown in Fig.~\ref{fig1}.
\begin{figure*}[ht]
  \centering
  \includegraphics[width = 2\columnwidth]{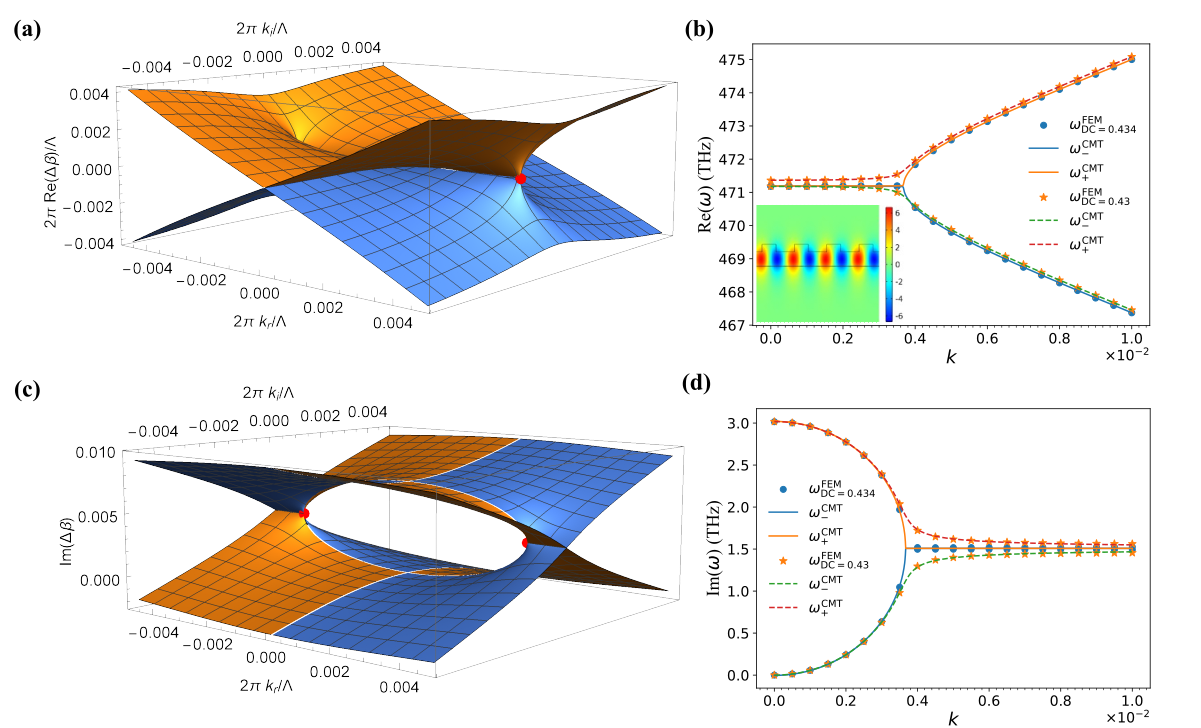}
    \caption
  {Nanophotonic gratings manifesting EPs. (a) Real part of eigenenergy surface variation with the complex k vector showing the branch cut along the real k axis with two EP lying along the ends of the branch cut (red dots). The wave vector and the eigenenergies here are made dimensionless as $k=2\pi k_{r}/\Lambda$ (b) eigenenergy variation along real k-axis,  the upper and lower bands are represented by orange and blue lines, respectively. This presentation highlights a distinctive parabolic division occurring at the exceptional point (EP) for a duty cycle set at 0.434 in contrast to star-marked data points for a non-EP condition (DC=0.43). The evident parabolic splitting is done using coupled mode theory and verified using Comsol (FEM simulations). Inset shows electric field distribution inside the grating at EP. (c) (c) Imaginary part of eigenenergy surface variation with complex k vector showing EP now existing at the end of the spectral energy gap. (d) Shows the imaginary eigenenergy plot along the real k-axis similar to (b). }
 \label{fig2}
\end{figure*}
We perform optimization of thicknesses of the various layers as well as the duty cycle via \textsc{comsol} simulations. \textcolor{black}{We found that an EP can be successfully attained by configuring the duty cycle (DC) and periodicity of the grating to 0.434 and 364 nm respectively, in conjunction with a silicon nitride thickness measuring 244.6 nm and an etch depth of 84.6 nm}. At this specific DC value, the forward and backward traveling waves radiate out of plane along the normal to the grating. 

To describe the physics of this setup we use the same approach as suggested in \cite{1072627,35721}, in which the net electric field is written as the summation of the perturbed (diffracted) electric field $\Delta E(x,z)$ arising from the corrugation of the grating, and the unperturbed field which is the fundamental mode passing through the uncorrugated waveguide. Thus, the total electric field can be written as:
\begin{equation}
    E(x,z)=\left[A_{+}(z)e^{iK_{0}z} + A_{-}(z)e^{-iK_{0}z}\right]\phi(x) + \Delta E(x,z) \nonumber
\label{eq:E_perturb}
\end{equation}
where $A_{+}(z),~A_{-}(z)$ are the forward and backward propagating modes traveling in-plane perpendicular to the grating ( $z$-axis) with wavevector $K_{0}=2\pi/\Lambda$, $\Lambda$ being period of the grating and $\phi(x)$ is spatial profile of the fundamental mode of waveguide in the transverse direction. One can expand the perturbation as a summation over partial waves to obtain differential equations governing the slow variation of amplitudes $A_{+}(z)$ and $A_{-}(z)$ as\cite{1072627}:
\begin{align}
    \left(\frac{\Delta\omega}{v_g} + i\frac{d}{dz}\right)A_{+}+ h_2 A_{-}+ i h_1 (A_{+}+A_{-}) &=0 \nonumber \\
    \left(\frac{\Delta\omega}{v_g} - i\frac{d}{dz}\right)A_{-}+ h_2 A_{+}+ i h_1 (A_{+}+A_{-}) &=0 \nonumber
\end{align}
This set of coupled equations describing the interaction between these two modes propagating in opposite directions can be expressed in a matrix form as:
\begin{align}
\label{eq2.1}
    i&\frac{d}{dz}\begin{bmatrix}
        A_+\\
        A_-
    \end{bmatrix} = \mathcal{H}\begin{bmatrix}
        A_+\\
        A_-
    \end{bmatrix} \nonumber\\ 
    \text{with}\nonumber\\ 
    \mathcal{H}&=\begin{pmatrix}
 -\Delta\omega/v_g\ - i h_1 &- h_2- i h_1 \\ 
 h_2+ i h_1 & \Delta\omega/v_g + i h_1
\end{pmatrix}\nonumber
\end{align}

\begin{figure*}[ht]
  \centering
  \includegraphics[width = 2\columnwidth]{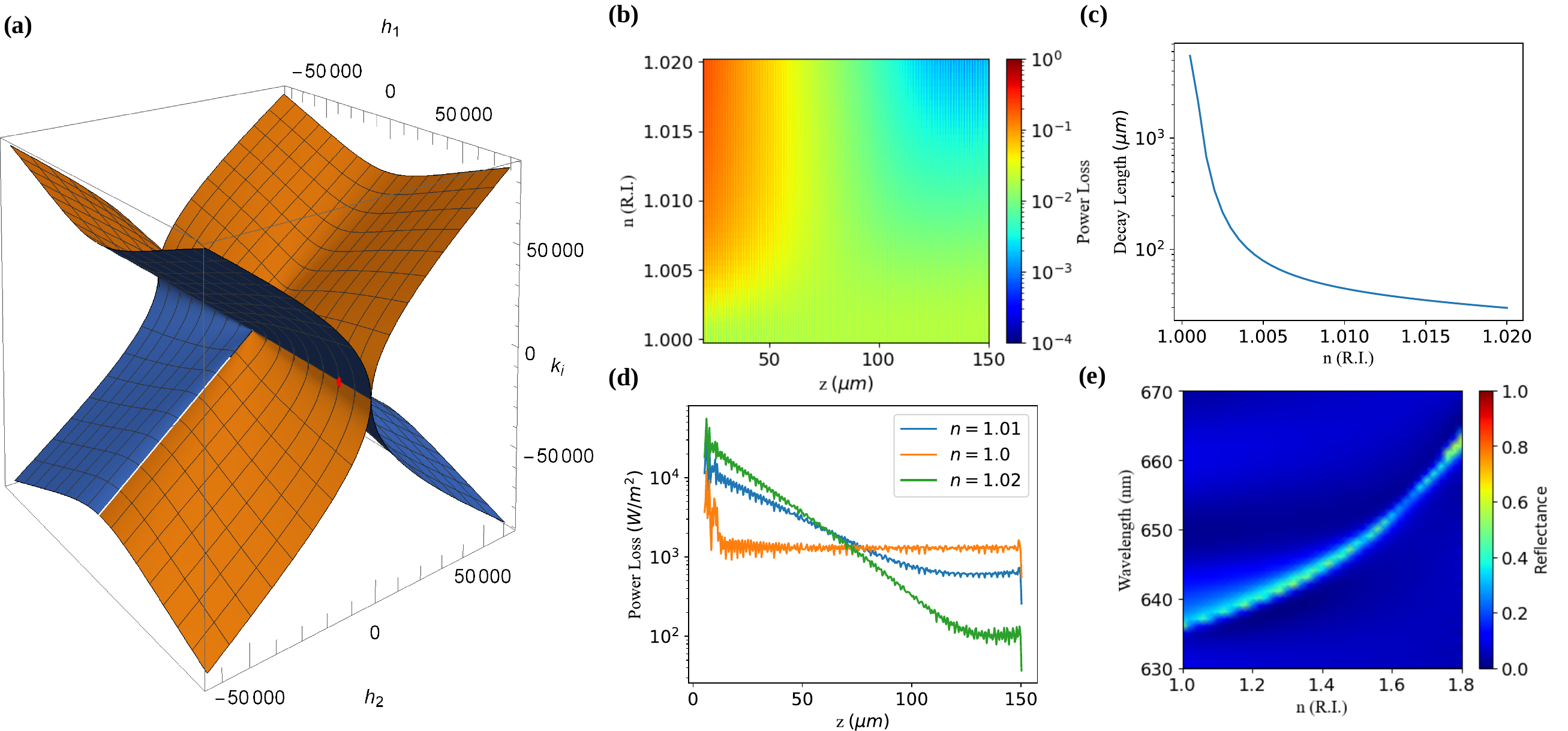}
   \caption{\textbf{Decay length based EP sensor:} (a) shows the variation of imaginary part (see supplementary for real part) of the complex wave vector with coupling coefficients $h_1$ and $h_2$ with exceptional points occurring along $h_2=0$ axis.  (b) Radiated power loss profile obtained from FEM simulations with along the grating for n = 1 and decay length variation (c) at $\lambda_{0}=\lambda^{EP}_{0}=636.1 \rm{nm}$ for changing refractive index demonstrating highest decay length for $n_{0}\rightarrow1$. (d) Intensity profile obtained from FEM simulations with various refractive indices. (e) Reflectance obtained along the grating showing the shift in the EP (having maximum intensity in the plot) by tuning the wavelength for different refractive indices.} 
  \label{fig3}
\end{figure*}
where $\Delta\omega$ is the detuning of laser frequency from the Bragg center frequency ($\omega_{0}=v_g K_0$), $v_g$ being group velocity of the medium, $h_1$ is the complex coupling coefficient of the propagating modes interacting via the radiative modes and $h_2$ is the coupling between the counter-propagating modes $A_{+}$and $A_{-}$. Both $h_1$ and $h_2$ are dependent on the refractive index $n_{\rm{eff}}$ of the medium present above the grating. This dependence will be exploited in the following sections. This equation is solved via the ansatz:
\begin{equation}
    A_{\pm}(z)= a_{\pm}~e^{ikz} + b_{\pm}~e^{-ikz}
\label{eq:AB_raw}
\end{equation}
where $a_{\pm}$and $b_{\pm}$ can be obtained from the boundary conditions $A_+(z=0)=1$ and $A_-(z=L)=0$ and wavevector given as:
\begin{equation}
    k=\pm \sqrt{( \Delta \omega/v_g +i h_1)^2-(h_2+i h_1)^2}
\label{eq:eigvals}
\end{equation}
From these expressions, it is clear that the above Hamiltonian has EP if the condition $Re(h_2 + i h_1)=0$ is satisfied \cite{Yulaev2022}, hence in our case the EP occurs when $h_2=0$\cite{Yulaev2022,1072627}. The results of these calculations are presented in Fig.~\ref{fig2}, where we plot the eigenenergy surface variation with complex wavector, showing the existence of EPs as the branch points. Cuts of this surface on the real wavevector are shown on the right in Fig.~\ref{fig2}, which shows the typical behaviour of the splitting of the of the real part of the frequency beyond a certain $k$ as we enter the PT symmetric state. The reverse behaviour is observed for the imaginary part as shown in the corresponding plot in Fig.~\ref{fig2}. In Fig.~\ref{fig2}(b) and (d) we also show the match between \textsc{comsol} simulation as well as coupled mode theory computations. Further, as the duty cycle is changed from the optimal point, the EP is shown to be destroyed.

\begin{figure*}[ht]
  \centering
  \includegraphics[width = 2\columnwidth]{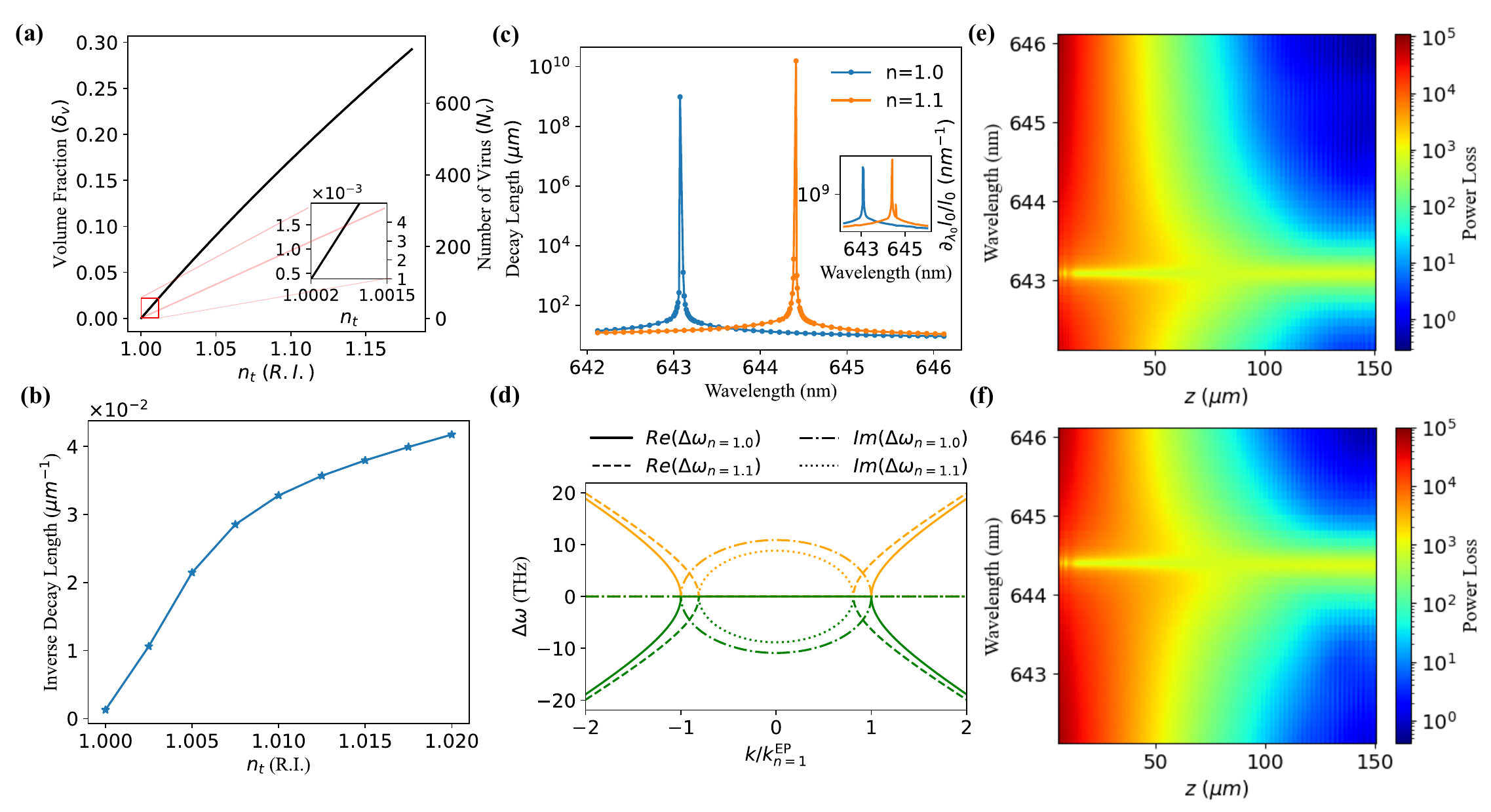}
  \caption{\textbf{EP sensing scheme for coronavirus detection: } (a) Dependence of refractive index of virus layer $n_t$ on the volume fraction $\delta_v$ and the corresponding number of viruses with variation in $n_t$. The inset shows the magnified picture for the same. We find that our proposed setup is capable to detect even 1 virus per grating period and a maximum of 750 viruses per grating period (b) Plot of inverse decay length showing a large change due to a small change in refractive index (using exponential fitting ). (c) Measurement of the decay length of the radiated intensity at a height of 4 $mu$m from the grating waveguides showing the highest quality factor and resolution. Inset shows the normalized decay length sensitivity (d) Effect of refractive index perturbation on band diagrams for $n = 1$ (all solid (real part) and dash-dotted (imaginary part) curves) and 1.1 (all dashed (real part) and dotted (imaginary part) curves). It shows that EP will not vanish but shift to some other k value with a change in refractive index. (e,f) Averaged radiation loss profile with z and wavelength for n=1.0 (e) and n=1.1 (f)}
\label{fig6}
\end{figure*}

\emph{EP sensing using optical decay length--}
In this section, we present a novel scheme of performing EP sensing by measuring the decay length of the radiated intensity from our architecture. The decay length $l_d$ is defined via the power profile, $P(z)=P(0)\exp{(-{z}/{l_d})}$. The decay length changes drastically with any small perturbation around EP. As shown in Fig.~\ref{fig3}(a), the EP occurs for $h_2=0$ whereas $h_1$ could be nonzero. Starting with this EP, optimized for refractive index $n=1$ for the medium above the grating, we look at the radiated power profiles as this refractive index changes in Fig.~\ref{fig3}(b). It is observed that the intensity profile sharply goes from a constant profile to a rapidly decaying profile as a function of position $z$ (in-plane direction, perpedicular to the grating). This behaviour holds for the complete range of refractive indices from $n=1.000$ to $n=1.020$. In Fig.~\ref{fig3}(d), we show the line plots for three representative refractive indices from Fig.~\ref{fig3}(b), where one can observe the variation of the power loss profile as the refractive index changes. From such measurements, one can in fact obtain a dependence of the decay length as a function of the refractive index as shown in Fig.~\ref{fig3}(c). Since the EP is optimized for a refractive index of 1.0, we see rapid variation of the decay length as the index changes from 1.0, suggesting that this quantity can be used for ultrasensitive refractive index detection. To understand this phenomenon, we consider an application to refractive index sensing, where using finite element simulations as well as CMT calculations we show that the decay length changes dramatically by three orders of magnitude by just changing refractive index from 1 to 1.002. This abrupt change in decay length can be used for extremely sensitive biochemical detection and we show how a single covid particle can be detected using this platform.

\emph{EP-based coronavirus sensor--}
In this segment, the design of the grating structure serves the purpose of functioning as a real-time detector for the coronavirus. To achieve this, subwavelength periodic gratings made of silicon nitride are functionalized with the human angiotensin-converting enzyme 2 protein (ACE2). This protein is specifically for binding the spike receptor-binding domain (RBD) protein of the virus and CoV-NL63\cite{Yan2020,Orenstein2009}. Notably, ACE2 acts as the cellular receptor for NL63, SARS-CoV, and SARS-CoV-2\cite{YANG2022131604}.

Subsequently, the configured grating system is positioned within the vicinity where the testing will take place. In this location, viruses will adhere to the ACE2 receptors, causing an instantaneous change in the effective refractive index of the medium situated above the grating. This alteration induces a drastic change in decay length. Thus, we are able to detect a single coronavirus particle within the test volume by observing a sudden change in the decay length. However, this detection is contingent upon the sensor being calibrated for the EP just by changing the excitation wavelength, thus being able to act as a real-time virus detector.

 The volume fraction relates to the refractive indices of air ($n_a$), effective virus layer ($n_t$), and the average refractive index of a single virus ($n_v$) as\cite{Landauer1978}:
\begin{equation}
\delta_v=\frac{\left(\frac{n_a^2 - n_t^2}{n_a^2 + n_t^2}\right)}{\left(\frac{n_a^2 - n_t^2}{n_a^2 + n_t^2}\right)-\left(\frac{n_v^2 - n_t^2}{n_v^2 + n_t^2}\right)}
\end{equation}
In Fig~\ref{fig6}, we calculate the allowed ranges of volume fraction of viruses binded on the ACE2 and the number of viruses using the $n_t$ and effective refractive index of all the medium above grating $n_{\rm{eff}}$. It is computed using the following relation \cite{YANG2022131604}:
\begin{align}
    n_{\rm{eff}}&=n_p-n_p e^{-\frac{2 d_p}{\delta}}+n_t e^{-\frac{2 d_p}{\delta}}-n_t e^{-\frac{2 d_p+2 d_t}{\delta}}\nonumber\\
    &+n_a e^{-\frac{2 d_p+2 d_t}{\delta}}
\label{eq:neff_nt}
\end{align}
where the grating depth $\delta$ is found by curve-fitting with the evanescent wave FEM simulation data plots as shown in supplementary.

We show that our proposed parameter of decay length enables extremely low FWHM in its spectra as shown in Fig.~\ref{fig6}(c) and discussed earlier. In the proposed scheme, the EP is present at a target refractive index for a certain initial concentration of coronavirus (refered hereafter as set point). For different such set points, virus number variation is shown in Fig~\ref{fig6}(a). In Fig~\ref{fig6}(b), for such a coronavirus sensor, we plot the inverse decay length as a function of the target refractive index in the presence of varying numbers of coronaviruses, showing high sensitivity. We further show in Fig.~\ref{fig6}(e,f) that EP wavelength is shifted for $n$ = 1 and 1.1 respectively. From Fig.~\ref{fig6}(d), it can be easily understood how EP can be maintained by tuning the laser wavelength in real time, although the refractive index might be changing due to increased number of coronaviruses accumulating on the chip. Hence, our sensing scheme enables us to employ the unprecedented sensitivity of EP physics without having to re-fabricate the chip once the virus concentration becomes too high.

\emph{Conclusion--} We have presented an EP based sensing scheme which relies on the measurement of spatial profile of the optical mode at a fixed wavelength as opposed to commonly used spectral measurement schemes. Our proposed platform is such that for large perturbation strengths, the EP is merely shifted in wavelength instead of being destroyed. This allows us to harness the unprecedented sensitivity of EP in real time measurements by just tuning the laser to the appropriate wavelength as the perturbation strength increases. We showed an application of this scheme to coronavirus sensing revealing a detection limit of one coronavirus within the test volume. Our findings will be important for the implementation of EP based sensing for low cost, real time measurements.

\emph{Acknowledgement--} P.S. and B.K. acknowledges support from Prime Minister's research fellowship (PMRF), Government of India. We acknowledge Prof. Venu Gopal Achanta for his invaluable insights. We acknowledge all LOQM members and Rahul Gupta for their knowledgeable discussions. A.K. acknowledges funding support from the Department of Science and Technology via the grant CRG/2022/001170.

\bibliography{ref}
\end{document}